\documentclass[12pt]{article}
\newcommand{\R}{{\rm I\kern-.2emR}}
\newcommand{\C}{{\rm \kern.25em\vrule height1.4ex
 depth-.12ex width.06em\kern-.31em C}}
\newcommand{\N}{{\rm I\kern-.16em N}}
\newcommand{\Z}{{\rm Z\kern-.35em Z}}
\newcommand{\bee}{\begin{equation}}
\newcommand{\ee}{\end{equation}}
\newcommand{\ba}{\begin{array}}
\newcommand{\ea}{\end{array}}
\newcommand{\bea}{\begin{eqnarray}}
\newcommand{\eea}{\end{eqnarray}}

\begin{document}
\thispagestyle{empty}
\begin{flushright}
MPI-PhT/95-71\\
AZPH-TH/95-17\\
July 1995
\end{flushright}
\bigskip\bigskip
\begin{center}
{\Huge{Super-Instantons, Perfect Actions, Finite Size Scaling
 and the Continuum Limit}}
\end{center}
\vskip 1.0truecm
\centerline{\bf
Adrian Patrascioiu}
\vskip5mm
\centerline{Physics Department, University of Arizona}
\centerline{Tucson, AZ 85721, U.S.A.}
\vskip5mm
\centerline{and}
\vskip5mm
\centerline{\bf Erhard Seiler}
\vskip5mm
\centerline{Max-Planck-Institut f\"{u}r
 Physik}
\centerline{ -- Werner-Heisenberg-Institut -- }
\centerline{F\"ohringer Ring 6, 80805 Munich, Germany}
\vskip 2cm
\bigskip \nopagebreak
\begin{abstract}
\noindent
We discuss some aspects of the continuum limit of some lattice models,
in particular the $2D$ $O(N)$ models. The continuum limit is taken
either in an infinite volume or in a box whose size is a fixed fraction
of the infinite volume correlation length. We point out that in this
limit the fluctuations of the lattice variables must be $O(1)$ and thus
restore the symmetry which may have been broken by the boundary
conditions (b.c.). This is true in particular for the so-called
super-instanton b.c. introduced earlier by us. This observation leads
to a criterion to assess how close a certain lattice simulation is to
the continuum limit and can be applied to uncover the true lattice
artefacts, present even in the so-called 'perfect actions'. It also
shows that David's recent claim that super-instanton b.c. require a
different renormalization must either be incorrect or an artefact
of perturbation theory.

\end{abstract}

\newpage\setcounter{page}1

\section{Introduction}
Lattice field theory can be considered as quantum field theory with a
cutoff. Of course the challenge is to dispose of the cutoff. From the
point of view of the lattice model that means letting the correlation
length become large. Combining this requirement with the desire of
working in a large thermodynamic box, one is quickly facing forbidding
costs in CPU time and memory. Although many techniques have been
proposed to circumvent this limitation, generally speaking they  fall into the
following two categories:

1. Choice of a better lattice action

2. Finite size scaling

Using the first technique one hopes that by complicating sufficiently the
lattice action the cutoff effects can be reduced so that continuum behavior
can be observed already at a correlation length of a few lattice units
with the perfect action of Hasenfratz and Niedermayer \cite{HN}.

With the
second technique one simulates the system in a box of finite
`physical size', i.e. a box whose
linear extent $L$ is a certain fraction of the thermodynamic (= infinite
volume) correlation length $\xi$. The idea is then to approach the
continuum limit by considering a sequence of lattices with fixed ratio
$z=L/\xi$ and extrapolating to the limit $\xi\to\infty$, using
certain assumptions about the asymptotic behavior. Thereby
it is believed that the so-called lattice artefacts can be eliminated.
A notable example of this philosophy is the work of the `Alpha
Collaboration' and its precursors \cite{Alpha}.
This procedure raises some questions, since the proposed form of the
approach to the limit does not have a solid theoretical basis
and different assumptions about it lead to quite different estimated
values of this limit \cite{PSprep}.

The crucial question is: do these techniques manage to reflect the true
continuum behavior up to some small corrections, or are they dominated by
lattice artefacts? One way to assess this lies (maybe surprisingly)
in studying the dependence of the data upon the boundary conditions (b.c.)
and possibly other constraints on individual spins.

In \cite{PS} we introduced `super-instanton b.c.' (s.i.b.c.) that are
characterized by fixing the spins at the boundary and in addition a spin
in the middle of the lattice. We pointed out that in the thermodynamic
limit one has to obtain the same results with s.i.b.c. as with more
conventional b.c.. In this paper we will show that the same is true
in the continuum limit even when it is taken in a box of `finite physical
size'. The practical use of this observation lies in the fact that one
can check to what extent this independence is fulfilled for particular
coupling parameters and box sizes. Although we do not present any numerical
data in this paper, our conclusion is that recent claims which appeared in
the literature regarding the continuum limit of $2D$ $O(N)$ models and
$4D$ gauge theories are unjustified in that they would not pass this test;
likewise is the claim that by employing perfect actions one can observe
continuum behavior already at small correlation length.

Our observation that super-instanton b.c. must lead to the same continuum
limit as say Dirichlet b.c. answers also David's \cite{David} recent
claim that they require a different renormalization: they cannot possibly
do so. Consequently, if in perturbation theory one finds, as he claims,
that in fact one does need additional renormalizations with s.i.b.c.,
then that is another proof that in these models perturbation theory fails
to produce the correct asymptotic expansion.

Before starting our discussion, we remind the reader of the general
procedure used to obtain a continuum limit of a lattice model: first one
has to find a point in parameter space where at least one dynamically
generated correlation length, called $\xi$, diverges. Then the continuum
correlation functions can be obtained by driving the system into this
critical point, using $\xi$ as the standard of length; calling the
lattice fields $s(x)$, this means that the $n$-point continuum correlation
function (Schwinger function) is given by
\bee
S_n(x_1,...x_n)=\lim_{\xi\to\infty}Z(\xi)^{-n/2}
\langle s(x_1\xi),...,s(x_n\xi)\rangle
\ee
where $Z(\xi)$ will be a suitably chosen field strength renormalization
constant. This will produce a massive continuum limit (of mass 1 with
the choice made in eq.(1)). Alternatively one can
construct a massless continuum limit by sitting right at a critical
point, introducing an arbitrary length standard $L_o$ that is sent to
$\infty$ and defining
\bee
S_n(x_1,...x_n)=\lim_{L_o\to\infty}Z(L_o)^{-n/2}
\langle s(x_1L_o) ...s(x_nL_o)\rangle
\ee

At least if the lattice fields $s(x)$ are bounded, it is unavoidable that the
field strength renormalizations $Z(\xi)^{-1/2}$ diverge for
$\xi\to\infty$, if the continuum limit is to be a quantum field theory
which has by necessity short distance singularities in its Schwinger
functions.

\section{Gaussian Computations}

To get a feeling for the situation, it is useful first to consider
free scalar fields $\Phi$ on the lattice $\Z^D$ with mass $m$ (including
the case $m=0$). First we look at the continuum limit on an infinite
lattice: the field is described by a Gaussian measure with covariance

\bee
C(x-y)=(-\Delta+m^2)^{-1}(x,y)
\ee
where  $\Delta$ is the lattice Laplacian. To obtain the continuum limit,
we have to drive the system into the critical point $m=0$, using $\xi=1/m$
as the standard of length. In other words, we study correlation functions
at distances that are fixed fractions of the correlation length $\xi=1/m$,
e.g.

\bee
\langle \Phi({x\over m})\Phi({y\over m})\rangle = C({x-y\over m})
\ee
and send $m\to 0$.

In dimension $D\ge 2$ this 2-point function is $O(m^{D-2})$, so to get a
nontrivial continuum limit in $D>2$ one has to introduce a divergent
field strength renormalization $Z(m)$ and define
\bee
\Phi_r(x)=\Phi({x\over m})Z(m)^{-1/2}
\ee
with $Z(m)=O(m^{D-2})$.
Then one obtains the continuum limit (for $x\neq y$)
\bee
\lim_{m\to 0} \langle \Phi_r(x) \Phi_r(y) \rangle ={1 \over (2\pi)^D}
\int d^Dp{ e^{ip\cdot (x-y)}\over p^2+1}
\ee
(The integral does not exist in the classical sense, but has to be
interpreted as follows: assuming without loss of generality that
$x_o-y_o\neq 0$, one integrates first over $p_o$ using the calculus
of residues; the remaining integral is then absolutely convergent.
To show convergence of the renormalized lattice two-point function to the
continuum limit, one uses the same trick. After the first integration has
been carried out, the dominated convergence theorem can be used.)

In $D=2$ no field strength renormalization is necessary.
But in all $D\ge 2$ we find for $x\neq y$
\bee
\lim_{m\to 0} \langle \bigl(\Phi_r(x)-\Phi_r(y)\bigr)^2\rangle =\infty,
\ee
which shows that the fluctuations of the renormalized fields diverge.
In $D\ge 3$ this divergence is due to the field strength renormalization,
whereas in $D=2$ it is due to the fact that $\lim_{m\to 0} C(0)=\infty$
because of the logarithmic infrared (IR) divergence.

It is easy to convince oneself that the fields $\Phi(x/m)$ and
$\Phi(y/m)$ become statistically independent in the continuum limit
for $x\neq y$: this is true for any b.c. and any $D\ge 2$ and is due to
the fact that $C({x-y\over m})/C(0)$ goes to zero.

Slightly less trivial is the case of an exponential of a free field
\bee
\Psi (x)=e^{iq\Phi(x)}
\ee
We limit ourselves to the study of $D=2$, because that is the most
interesting case, and for $D>2$ the continuum fields would become
non-tempered, i.e. the correlation functions would develop exponential
singularities.
Even though the fields $\Phi$ did not require renormalization in $2D$,
their exponentials do, as can be found essentially already in
Coleman's paper \cite{Cole}. If we define
\bee
\Psi_r(x)=Z(m)^{-1/2}\Psi(x/m)
\ee
with $Z(m)=m^{q^2/2\pi}$, the correlation functions of the renormalized
fields $\Psi_r$ will have nontrivial continuum limits. But it is
interesting to look at the continuum limit from the point of view of
the lattice fields $\Psi(x/m)$: then we find restoration of the $O(2)$
symmetry in accordance with
the Mermin-Wagner theorem \cite{MW} in the limit $m\to 0$. Explicitly
\bee
\langle \Psi({x\over m} \Psi({y\over m})^\ast \rangle =
\exp\bigl(-q^2(C(0)-C({x-y\over m}))\bigr)
\ee
which goes to 0 as $m\to 0$ because $C(0)=O(|\ln(m)|)$.

One can also establish that the fields $\Psi(x/m),\Psi(y/m)$ become
statistically independent in the limit $m\to 0$ for $x\neq y$.
Furthermore each field $\Psi(x/m)$ will be distributed uniformly on
the unit circle in this limit. To see this, it suffices to consider
\bee
\langle \Psi(x/m)^{n_x}\Psi(y/m)^{n_y} \rangle=
e^{-{q^2\over 2}\bigl[C(0)(n_x^2+n_y^2)+2n_xn_yC({x-y\over m})\bigr]}
\ee
for $n_x,n_y\in\Z$.
It is easy to see that for $n_x^2+n_y^2\neq 0$ this goes to zero,
while for $n_x=0=n_y$ it is equal to $1$. Thus we have
\bee
\lim_{m\to 0}\langle \Psi(x/m)^{n_x}\Psi(y/m)^{n_y}\rangle=
\delta_{n_x0}\delta_{n_y0}
\ee
from which the claim follows.

Next we turn to the continuum limit in a box. The linear extent $L$
is to be kept fixed in `physical units', i.e. we choose $L=l/m$ with
$l$ fixed. We may use Dirichlet, periodic or any other classical b.c..
The discussion of the continuum limit proceeds as above and the equations
are changed only by replacing $C(x-y)$ with $C_{bc}(x,y)$,
the covariance with the appropriate b.c.; we find as above:

(1) In $D>2$ the Gaussian field $\Phi$ requires a wave function
renormalization, leading to divergent fluctuations of the massive free
field in the continuum limit.

(2) In $D=2$, $\Phi$ requires no wave function renormalization,
but the fluctuations of the free massive field diverge in the continuum
limit due to the IR divergence of $C_{bc}(x,x)$.

(3) In $D=2$ the renormalized exponentials $\Psi_r(x)$ of the free
field require a wave function renormalization; their expectation values
are not $O(2)$ symmetric in accordance with the symmtry breaking mass term.

(4) In $D=2$ the unrenormalized exponentials $\Psi(x)$ show restoration
of the $O(2)$ symmetry, because from the point of view of the lattice
(i.e. measured in lattice units) the box is becoming infinitely large
and the symmetry breaking by the mass term disappears in the limit.
The fields $\Psi(x_i/m)$ become statistically independent for different
$x_i$.

Next we turn to s.i.b.c.. They are defined as 0 Dirichlet b.c. at the
boundary of the box, at distance $O(1/m)$ from the origin,
together with the constraint $\Phi(x_c)=0$ where
$x_c$ is a point in the `middle of the lattice', e.g. the origin.
The Green's function $\langle\Phi(x)\Phi(y)\rangle$
with these b.c. can be expressed
in terms of the Dirichlet Green's function $C_D$ as follows
\cite{PS, Sok}:
\bee
\langle \Phi(x)\Phi(y)\rangle_{s.i.b.c.}
=C_D(x,y)-{C_D(x,x_c) C_D(x_c,y)\over C_D(x_c,x_c)}
\ee
More generally we may require that $\Phi(x_c)=a$, whereas at the edges
of the box we still have 0 Dirichlet b.c.. In this case the two-point
function becomes
\bee
\langle \Phi(x)\Phi(y)\rangle_{s.i.b.c.}=C_D(x,y)-
{C_D(x,x_c) C_D(x_c,y)\over C_D(x_c,x_c)}
+a^2{C_D(x,x_c) C_D(x_c,y)\over C_D(x_c,x_c)^2}
\ee
and there is also a nonvanishing one-point function
\bee
\langle \Phi(x)\rangle_{s.i.b.c.}=a{C_D(x_c,x)\over C_D(x_c,x_c)}
\ee

Looking at eqs.(13),(14),(15) one sees at once that in $2D$, if we replace
$x$ by $x/m$ etc., the extra terms go to zero because their denominators
blow up as $m\to 0$. In $D>2$, after field strength renormalization,
and replacing $x$ by $x/m$ etc, the extra terms go to zero because
the numerators do not have enough renormalization factors.
It should not come as a surprise that the additional constraint
$\Phi(x_c)=a$ does not leave any trace in the continuum limit
of the renormalized fields, because in the continuum it is impossible to
impose a Dirichlet condition at a point in $D>1$
(or more generally on a set of zero capacity).

It is clear from eqs.(8) and (9) that the situtation is analogous
for the exponential fields $\Psi(x/m)$ and $\Psi_r(x)$.

There is another continuum limit that can be discussed: We can put the
massless Gaussian field in a box of size $L$, with 0 Dirichlet b.c.
to avoid trouble from the zero mode, and take $L$ as the standard of
length. The continuum limit is now defined by the limit $L\to\infty$
of the correlations of the renormalized fields
\bee
\Phi_r(x)=Z(L)^{-1/2}\Phi(xL)
\ee
with $Z(L)=L^{2-D}\quad (D\ge 2)$, or (only in $2D$):
\bee
\Psi_r(x)=Z(L)^{-1/2}\Psi(xL)\equiv Z(L)^{-1/2}e^{iq\Phi(xL)}
\ee
with $Z(L)=L^{-q^2/2\pi}$.

Also in this massless case we can consider s.i.b.c.. The results are
analogous to the massive case discussed above:

(1) The renormalized fields $\Phi_r$ show divergent fluctuations as in
eq.(5) in the continuum limit.

(2) The renormalized fields $\Psi_r$  ($D=2)$ show no $O(2)$ symmetry.

(3) The lattice fields $\Psi$ show restoration of the $O(2)$ symmetry.

(4) S.i.b.c. become identical to Dirichlet b.c. in the continuum limit.

\section{$2D$ $O(N)$ Models}

In this section we want to show that what we found for the Gaussian
models also holds more generally, in particular for the $2D$ $O(N)$
models. These models describe configurations of classical spins
$\{s(x)\}, s(x)\in\R^N, s(x)=1, x\in\Lambda$, where $\Lambda$ is the
lattice $\Z^D$ or a finite part of it (like a box of size $L$).
For definiteness we may consider the standard nearest neighbor action
(s.n.n.a.)
\bee
S=\sum_{\langle x y\rangle} s(x)\cdot s(y)
\ee
(even though that is inessential) and the Gibbs state
induced by it via the Boltzmann factor $\exp(-\beta S)$.

First let us discuss the $xy$-model ($N=2$). This model has the famous
Kosterlitz-Thouless transition from a massive phase at $\beta<\beta_{crt}$
to a massless one at $\beta\ge\beta_{crt}$ \cite{KT,FS}.
A massive continuum limit is constructed by driving
$\beta\to\beta_{crt}$
from below, using the correlation length $\xi=1/m$ as the unit of length,
as in the Gaussian models above. For instance the two-point Schwinger
function becomes
\bee
S_2(x,y)=\lim_{\beta\to\beta_{crt}-0} Z(\beta)^{-1}
\langle (s(x\xi)\cdot s(y\xi))\rangle
\ee
First let us point out that there has to be a field strength
renormalization $Z(\beta)^{-1}$ that diverges for $\beta\to\beta_{crt}-0$,
to compensate for the fact that without it the two-point function would
go to zero. This can be seen as follows:
Let $\Lambda_{r\xi}$ be a box of size $r\xi$ within the infinite lattice
$\Z^D$. Then the root mean square (rms) magnetization in that box is
given by
\bee
M_{rms}={1\over (r\xi)^2} \sqrt{\langle
\bigl(\sum_{x\in\Lambda_{r\xi}}s(x)\bigr)^2\rangle}.
\ee
Assuming the two-point function is everywhere nonnegative -- a
well known though not rigorously proven fact -- this is bounded by
$\sqrt{(\chi/(r\xi)^2}$  (remember that there is no subtraction of
a disconnected contribution because the $O(2)$ symmetry is unbroken).
Now it is well known that
\bee
{\chi\over\xi^2}=O(\xi^{-\eta})
\ee
for $\beta\to\beta_{crt}-0$, and according to the Kosterlitz-Thouless
theory \cite{KT} $\eta=1/4$ (reasonably well confirmed by the numerical
simulations \cite{Gupta,Hasb}). This implies that the rms
magnetization over a box of size $r\xi$ goes to 0 in the continuum limit.
But the $M_{rms}^2$ is nothing but a double average of the two-point
functions over that box, and under the positivity assumption made above
it follows that
\bee
\lim_{\beta\to\beta_{crt}-0}\langle (s(x\xi)\cdot s(y\xi))\rangle=0
\ee
for $x\neq y$.
On the way we have learned that the correlation between two spins
located at a distance $x\xi$ will go to zero as $\xi\to\infty$,
contrary to a widespread belief (this is also in agreement with the
Ornstein-Zernike behavior as discussed below).

It should be pointed out that the existence of a nontrivial continuum
limit requires that the same field strength renormalization that is needed
for the two-point function also works for the higher $n$-point functions.
Composite fields (products of fields a the same point) will require
field strength renormalization as well (cf. eq.(11)). Conversely, since
the field strength renormalization has to diverge in the continuum limit,
all the correlations of the unrenormalized lattice spins at `physical'
distances will go to zero in that limit, and those spins will become
statistically independent.

Next let us turn to the continuum limit in a box $\Lambda_{r\xi}$
of size $r\xi$. As before we find that the rms magnetization is given by
\bee
M_{rms}={1\over (r\xi)^2} \sqrt{\langle
(\sum_{x\in\Lambda_{r\xi})_{r\xi}}s(x))^2\rangle}={\chi_{r\xi}
\over\xi^2}
\ee
Note that we denote by $\xi$ the infinite volume correlation length
and by $\chi_{r\xi}$ the susceptibility in the finite box. We invoke now
the hypothesis of finite size scaling (FSS) \cite{FSS},
which says that
\bee
\lim_{\beta\to\beta_{crt}-0}{\chi_{r\xi}\over\chi_\infty}=f(r)
\ee
to conclude as before that also the two-point function in the box will
go to zero in the limit $\xi\to\infty$.

As in the Gaussian models, one can also discuss a continuum limit
in the massless (KT) phase. Since there is no mass to set the scale,
one chooses an arbitrary diverging scale unit $L_o$ and considers the
limit
\bee
S_2(x,y)=\lim_{L_o\to\infty} Z(L_o)^{-1}
\langle (s(xL_o)\cdot s(yL_o))\rangle
\ee
Note that $\beta\ge\beta_{crt}$ is kept fixed in this limit.
It is rigorously known \cite{FS,FFS} that $\eta>0$ (KT theory predicts
in fact $1/4\ge\eta>0$), so we can conclude as
above that the spin-spin correlation without field strength
renormalization will vanish in the continuum limit
and a similar argument can be made for the continuum limit in a finite
box of sixe $L_o$.
It should be stated (though it may be obvious) that the $O(2)$
symmetry of the unrenormalized spins is restored in the limit,
no matter what symmetry breaking b.c. we used. This is of course
simply a consequence of the Mermin-Wagner theorem \cite{MW} which
says that on the infinite $2D$ lattice, a continuous symmetry cannot be
broken spontaneously.

Now we are ready to discuss s.i.b.c. for the $O(2)$ model. Again they are
defined by fixing a spin in the middle in addition to imposing Dirichlet
(fixed) b.c. at the boundary of our box. Because the spin in the middle
becomes uncorrelated with all the spins that have a distance $O(\xi)$
or $O(L_o)$, respectively, in both limits (massive and massless)
discussed above, s.i.b.c. become equivalent to Dirichlet b.c..

Let us now extend the discussion to the $O(N)$ models with $N>2$.
Elsewhere \cite{Pat,Lat92} we have presented arguments for the existence
of a finite $\beta_{crt}$ such that for $\beta\ge\beta_{crt}$ there is a
massless phase in all these models. Accepting this point of view,
the discussion can be taken over unchanged from the $O(2)$ model.
The conventional wisdom -- with which we disagree --
states, however, that the model is critical only at $\beta=\infty$.
We want to point out that even if we accept this point of view for the
sake of the argument, the same conclusions as before hold.

Since by assumption there is no massless phase, we only have to discuss
the massive continuum limit. According to the conventional wisdom
the correlation length $\xi$ and the magnetic susceptibility $\chi$
behave as follows for $\beta\to\infty$ \cite{BZ}:
\bee
\xi\propto\beta^{-{1\over N-2}}e^{{2\pi\beta\over N-2}}
\ee
\bee
\chi\propto\beta^{-{N+1\over N-2}}e^{4\pi\beta\over N-2}
\ee
which would imply
\bee
{\chi\over \xi^2}=O(\beta^{-{N-1\over N-2}})
\ee

Since this vanishes in the limit $\beta\to\infty$, we obtain again
the conclusion that in the limit $\xi\to \infty$ the spin-spin
correlations at distances that are fixed multiples of the correlation
length will vanish and the system, from the point of view of the
lattice spins, restores the $O(N)$ symmetry in that limit.

Quite generally, and independently of the question whether the
conventional scenario is true or false, the existence of a
continuum limit describing a quantum field theory enforces a divergent
field strength renormalization, and as above we conclude that therefore
the lattice spins at distances proportional to the correlation length
will become statistically independent of each other.

Tree level perturbation theory (which is uncontested) reveals that the
spins stay well ordered only over a distance $O(\xi_{PT})$ where
\bee
\xi_{PT}=e^{2\pi\beta\over N-1}
\ee
and this is becoming arbitrarily small with respect to the correlation
length $\xi$, whether we subscribe to the conventional scenario (eq.(20))
or believe in the existence of a critical point at finite $\beta$.

Likewise it follows that s.i.b.c. are equivalent to Dirichlet b.c. in the
continuum, and it is an obvious generalization that fixing any finite
number of spins will have as little effect as fixing one, i.e. no effect
in the continuum.

The necessity of a divergent field strength renormalization
(and the absence of spontaneous symmetry breaking at any $\beta$,
which in $2D$ follows from the Mermin-Wagner theorem),
is also in full accordance with the so-called Ornstein-Zernike behavior
of correlations at large distance (see for instance \cite{PL});
one can even obtain a prediction for the behavior of the field strength
renormalization from the Ornstein-Zernike behavior.
The Ornstein-Zernike postulate, which has been proven in some cases like
the Ising model \cite{PL}, but is expected to hold generally in massive
models, since it corresponds to the requirement that the model
describes massive particles with an isolated mass shell, says:
\bee
\langle s(x\xi)\cdot s(y\xi)\rangle \cong \xi^{2-D-\eta}
{|x-y|}^{-(D-1)/2} \exp(-|x-y|/\xi)
\ee
for $|x|>>\xi$.
It can be seen immediately that this expression times a field strength
renormalization factor $Z^{-1}$ has a continuum limit if and only if
\bee
Z=O(\xi^{2-D-\eta})
\ee
The Gaussian models discussed in the previous section have $\eta=0$; so
eq.(29) generalizes the result found there. From the so-called infrared
bounds it follows that $\eta\ge 0$ (see for instance \cite {FFS}),
an inequality that is also required if the continuum
theory is to be Osterwalder-Schrader positive; of course logarithmic
corrections to the pure power behavior assumed in eq.(29) are legitimate,
provided they correspond to a stronger divergence of $Z$ than in the
free Gaussian model. In all these cases the Ornstein-Zernike behavior
leads to the same conclusions as our other considerations.

\section{Discussion and Conclusions}

We have seen that fixing a finite number of spins has no effect on the
continuum limit in a box of finite physical size or in the infinite volume.
This is to be contrasted with the claim made by David \cite{David} in a
comment to our papers \cite{PS, PSG}. David claimed, that fixing a spin
at the origin in addition to Dirichlet b.c. (imposing s.i.b.c.)
will necessitate extra renormalizations; as we have seen here,
there is no effect of the extra spin on the continuum limit. So if in fact
in perturbation theory one finds the need for such an extra renormalization,
then this is just another proof that perturbation theory does not produce
the correct asymptotic expansion in these models.

Furthermore we have learned that in the continuum limit the system gets
disordered on the scale of the correlation length, in the sense that spins
located at a fixed finite fraction of the correlation length will become
decorrelated and this phenomenon even occurs in a box of finite physical
size.
This is in accordance with the properties of continuum quantum fields
known from axiomatic quantum field theory. We have learned there long ago
that continuum quantum fields and their correlations are never functions,
but have distributional character, in other words there are large
fluctuations at short continuum distances. So if the phenomena found here
did not occur, there would be no chance to have a continuum limit
satisfying e.g. the Osterwalder-Schrader axioms \cite{OS}.

In \cite{PS, PSG} we pointed out the importance of certain `defects'
dubbed super-instantons in disordering the $O(N)$ models at large $\beta$.
These are configurations that turn the spin gradually from a certain
value in the center of a region to a different one at its edge; we
stressed that these configurations require arbitrarily little energy
and have entropy corresponding to their position and scale;
therefore they should be abundantly present even at low temperature.
Here we found that at the scale of the correlation length, constraining
the spin at the center has no effect. This is also a manifestation of the
fact that super-instantons become so abundant in the continuum limit that
they disorder the system even at the scale of the correlation length, and
thus forcing an extra super-instanton into the system has no effect.

There is a useful lesson to be drawn from our observations: Since we
now know that in the continuum limit fixing a finite number of spins
cannot have any effect, by doing precisely this and checking how much
the physics changes, we can assess how close our results are to the true
continuum limit. This is of relevance in particular in studies where
small lattices are used to extract information about the presumed
continuum limit; notable examples are the work of the `Alpha
collaboration' \cite{Alpha} studying the running coupling, the
work by Kim \cite{Kim} and by Caracciolo et al \cite{Sokal} on finite
size scaling in the $O(3)$ model and various claims which have appeared
in the literature regarding the miraculuous properties of the
improved/perfect actions in simulating continuum physics already at rather
small correlation lenghts \cite{HN,Lep}. We intend to return to this question
in a more quantitative fashion elsewhere \cite{PSprep}; however we would
like to state here that this type of claims, that from lattices of modest
size one can learn true continuum behavior, are in our opinion false:
whatever the action may be, the lattice must be large enough to allow the
typical configuration to resemble a gas of super-instantons, i.e. to
restore a certain symmetry required by the Gibbs measure.

This remark applies equally to gauge theories. Contrary to the claim of
the `Alpha Collaboration' and its precursors, we believe that their data
on the running of $\alpha_s(Q)$ do not reveal the true continuum behavior
of QCD. Indeed for example the study of this running in $SU(2)$
\cite{Alpha} involves lattices with $L\le 20$ and $\beta\ge 3$, a regime
in which the typical configuration corresponds to small fluctuations
around a well ordered state, rather than a gas of super-instantons;
this explains also the excellent agreement they found in the running of
$\alpha_s(Q)$ with the prediction of perturbation theory.

\end{document}